\documentclass[showpacs,floatfix,preprint,amsmath,amssymb]{revtex4}
\usepackage{bm}
\usepackage{epsfig}
\usepackage{graphicx}
\usepackage{amsmath}
\usepackage{dcolumn}
\usepackage{epstopdf}

\begin{document}

\title{A three-body calculation of incoherent $\pi^0$ photoproduction on a deuteron}

\author{
A.~Fix$^{1}$\footnote{\emph{eMail address:} fix@tpu.ru} and
H.~Arenh\"ovel$^{2}$\footnote{\emph{eMail address:} arenhoev@uni-mainz.de}}
\affiliation{\mbox{$^1$Tomsk Polytechnic University, 534050 Tomsk, Russia}
\\ \mbox{$^2$Institut f\"ur Kernphysik, Johannes
  Gutenberg-Universit\"at Mainz, 55099 Mainz, Germany}} 

\date{\today}

\begin{abstract}
Incoherent $\pi^0$ photoproduction on a deuteron  in the
$\Delta(1232)$ region is treated in a three-body scattering approach
using separable two-body
interactions. Results are presented for total and differential
cross sections. It turns out that the role of higher order terms
beyond the first order in the multiple scattering series is
insignificant, and their inclusion cannot explain the existing
discrepancy between theory and experiment.
\end{abstract}


\maketitle

\section{Introduction}
The role of the final state interaction (FSI) in incoherent $\pi^0$
photoproduction on a deuteron
\begin{equation}\label{eq1}
\gamma + d \to \pi^0 + n + p
\end{equation}
has been studied by various groups
\cite{Laget,Darwish,FixArenh,Levchuk,Tarasov,Nakamura}. In general, the
theoretical treatment of this reaction was based on the multiple
scattering picture, in which, however, only the first order terms with
respect to the final $NN$ and $\pi N$ interactions are taken into
account. According to these studies, the main FSI effect arises from
$NN$ rescattering,
whereas the contribution from $\pi N$ rescattering is rather small.

It is well known that in the reaction in Eq.~(\ref{eq1}) the
first-order inclusion of the $NN$ interaction has a particularly
strong effect compared to the impulse approximation (IA). The reason for
this feature is the fact, that in contrast to processes with charged
pions, $\gamma d\to \pi^+nn/\pi^- pp$, for the
incoherent $\pi^0$ production (\ref{eq1})
the impulse approximation contains a spurious contribution
of the coherent reaction ($\gamma d\to \pi^0d$) since the final plane
wave is not orthogonal
to the deuteron ground state~\cite{Noble,Cannata,FixArenh}. Indeed,
projecting out the ground state from the final plane wave, the
socalled modified IA, comprises already the dominant part of the
first order FSI correction~\cite{FixArenh}. The
remaining FSI effect is of the same order as for charged pion
production. Further incorporation of $\pi N$
rescattering gives an additional (however much less significant)
decrease. Thus the total first-order FSI effect in the
$\Delta$-resonance region is a decrease of the total cross
section by about 30~$\%$ compared to the one predicted by the
pure spectator model (IA).

On the other hand, the calculation including only the first order
rescattering terms still overestimates the experimental total cross
section~\cite{Krusche,Siodlaczek} by about  15 $\%$ at the
$\Delta(1232)$ peak. { It appears reasonable to assume that the
remaining difference could be assigned to the neglect of the higher
order terms, which can cause an additional broadening of the $\Delta$
resonance and, consequently, can lead to a lowering of the $\Delta$
peak in the cross section. 

In the present work we study the role of the higher orders of the
multiple scattering series in the reaction (\ref{eq1}).}  To this end,
we calculate the reaction amplitude using the three-body scattering
theory. In the next section we briefly outline the formalism. Our
approach is based on a separable representation of the driving
two-body $\pi N$ and $NN$ interactions. As is well known, in this case
the original three-body equations simplify to a set of equations of
Lippman-Schwinger type for a system of coupled quasi-two-body channels.
To reduce this set into an easily solvable one-dimensional form, we
apply an expansion into partial waves. In Sect.~\ref{SectResults} we present
our results and compare them with existing experimental data. We also
discuss the importance of the multiple scattering corrections.
Conclusions are given in the final Section~\ref{Conclusion}.

\section{Formalism}

In the present approach we use for the description of the final $\pi^0np$
three-body state two coupled two-body channels, each consisting of a
quasiparticle formed by two of the three particles and the remaining
one as a spectator. Thus each quasiparticle is an interacting
two-body system. The two channels are in detail: 

(i) Channel ``$d$'' consisting of a deuteron as a quasiparticle
  with two interacting nucleons and a pion as spectator. 

(ii) Channel ``$\Delta$'' consisting of an interacting
  nucleon-pion system forming a $\Delta$ as quasiparticle and a
  spectator nucleon. 

In the following we use $\alpha,\,\beta,\dots\in\{d,\Delta\}$ to
label the channels and the corresponding quasiparticles, while
$a,\,b,\dots$ are used for the corresponding spectators. In this
notation the channel $\alpha$ consists of a spectator $a$ and two
interacting particles $(bc)$ forming the quasiparticle $\alpha$.

\begin{figure}[h]
\begin{center}
\resizebox{0.95\textwidth}{!}{%
\includegraphics{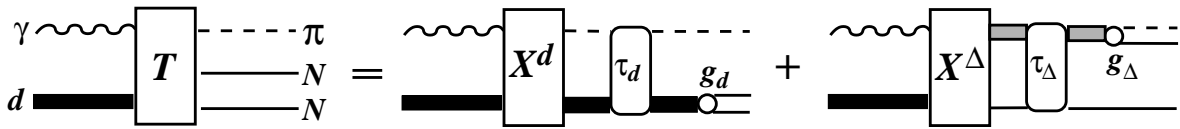}}
\caption{Diagrammatic representation of the $T$-matrix in Eq.~(\ref{T-matrix}). }
\label{fig1a}
\end{center}
\end{figure}
Treating the electromagnetic interaction perturbatively in lowest
order one obtains for the reaction $T$-matrix
\begin{equation}\label{T-matrix}
T=\sum_{\alpha\in\{d,\Delta\}} X^\alpha\,\tau_\alpha g_\alpha\,,
\end{equation}
where $X^\alpha$ denotes a channel amplitude, $\tau_\alpha$ the
channel propagator and $g_\alpha$ the quasiparticle vertex for
$(bc)\to b+c$. A graphical representation of the $T$-matrix in terms
of the channel amplitudes is shown in Fig.~\ref{fig1a}.

\begin{figure}[ht]
\begin{center}
\resizebox{0.95\textwidth}{!}{%
\includegraphics{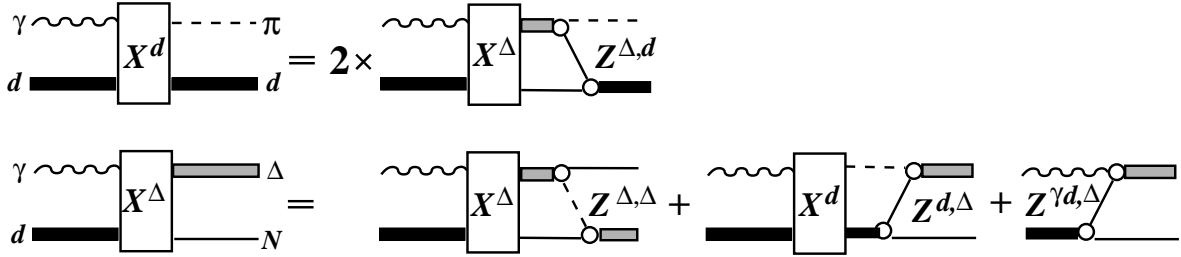}}
\caption{Diagrammatic representation of the system of three-body equations of
  Eq.~(\ref{Eq1_1}). The factor 2 in the first equation arises from
  the  symmetrization of the two nucleons.}
\label{fig1}
\end{center}
\end{figure}

The amplitudes $X^\alpha$ obey a set of coupled equations, which can
be derived from the Faddeev three-body formalism under the assumption
of separable two-body interactions. In operator  form  they read
\begin{equation}\label{Eq1_1}
X^\alpha=
\sum_{\beta\in\{d,\Delta\}}X^\beta\,\tau_\beta\,Z^{\beta,\alpha} 
+ Z^{\gamma d,\alpha}\,,\quad \alpha\in\{d,\Delta\} 
\,.
\end{equation}
These equations are shown in a graphical representation in Fig.~\ref{fig1}.

The driving terms $Z^{\alpha,\beta}$ describe the exchange
of a particle $c$ between the quasiparticles $\alpha$ and $\beta$.
The term $Z^{\gamma d,\Delta}$, forming the inhomogeneous part of the
set (\ref{Eq1_1}), contains the electromagnetic vertex $\gamma
N\to\Delta$. Obviously one has $Z^{\gamma d,d}=0$.

In momentum space the potentials $Z^{\alpha,\beta}$ have the following
form
\begin{equation}\label{Eq1_2}
Z^{\alpha,\beta}(\vec{p}_a,\vec{p}_b;W)=
\frac{g_\alpha(q_{\alpha})g_\beta(q_{\beta})}
{W-E_a(p_a)-E_b(p_b)-E_c(|\vec{p}_a+\vec{p}_b|)+i\epsilon}\,.
\end{equation}
Here $W$ denotes the total energy in the center-of-mass (c.m.) frame,
$\vec{p}_a$ and $\vec{p}_b$ the c.m.-momenta of the spectator
particles of the channels $\alpha$ and $\beta$, respectively. The
relative momenta $\vec{q}_{\alpha/\beta}$ between the spectators of
the channels $\beta /\alpha$, respectively, and the exchanged particle
in the arguments of the vertices $g_{\alpha/\beta}(q_{\alpha/\beta})$
are treated nonrelativistically, e.g. 
\begin{equation}\label{Eq1_3}
\vec{q}_{\alpha}=\vec{p}_b+\frac{M_b}{M_b+M_c}\vec{p}_a\,,
\end{equation}
with $M_c$ denoting the mass of the exchanged particle, whereas for
the particle energies we use the relativistic relation
$E_a(p)=\sqrt{p^2+M_a^2}$.

To reduce Eq.~(\ref{Eq1_1}) to a numerically manageable form
we exploit a partial wave expansion of the amplitudes $X^{\alpha}$ in
terms of the total angular momentum $J$ and the isospin $T$. We use
the $LS$ coupling scheme by coupling the total angular momentum
$\vec{j}_\alpha$ of the quasiparticle with the spin
$\vec{s}_a$ of the third particle to the total channel spin
$\vec{S}_\alpha$. The orbital momentum $\vec{L}_\alpha$
is then coupled with $\vec{S}_\alpha$ to the total angular momentum
$\vec{J}$.
For the given values of photon
polarization $\vec{\epsilon}_\lambda$, initial deuteron spin
projection $M_d$, total spin $S_\alpha$ with projection $
M_{S_\alpha}$, and total isospin $T$ of
the final quasi-two-body state~$\alpha$ the partial wave expansion of
the channel amplitudes  $X^{\alpha,T}_{\lambda M_d\,S_\alpha M_{S_\alpha}}
(\vec{k},\vec{p};W)$ reads
\begin{eqnarray}\label{Eq1_16}
X^{\alpha, T}_{\lambda M_d\,S_\alpha M_{S_\alpha}}(k,\vec{p}\,;W)
&=&\frac{N_d}{\sqrt{4\pi}}
\sum\limits_{J^\pi}\sum\limits_{L,S,M_S}\sqrt{2L+1}\,\sum\limits_{L_\alpha, M_{\alpha}}
X^{\alpha, T;J^\pi}_{LS,L_\alpha  S_\alpha}(k,p;W)\,Y^*_{L_\alpha
    M_{\alpha}}(\hat{p})
\nonumber\\
&&\times(1\lambda\,1M_d|SM_S)
(L0\,SM_S|JM_S)(L_\alpha M_{\alpha}\,S_\alpha M_{S_\alpha}|JM_S)\,,
\end{eqnarray}
where the factor $N_d$ takes  into account the deuteron
normalization. In Eq.~(\ref{Eq1_16}) the $z$ axis is chosen along the
initial photon momentum $\vec{k}$.

With the help of this  partial wave decomposition one obtains from
Eq.~(\ref{Eq1_1}) a set of one-dimensional coupled equations for each value of
$J$, parity $\pi$ and isospin $T$ in the following form (for simplicity we
drop the energy $W$ in the arguments)
\begin{eqnarray}\label{Eq1_4}
X^{\alpha, T;J^\pi}_{LS,L_\alpha S_\alpha}(k,p)&=&
Z^{\gamma d,\alpha, T;J^\pi}_{LS,L_\alpha S_\alpha}(k,p)\nonumber\\
&+&
\sum\limits_{\beta\in\{d,\Delta\}}\sum\limits_{L_\beta, S_\beta }
\int \frac{p^{\prime\,2}dp^{\prime}}{(2\pi)^3}\,
X^{\beta, T;J^\pi}_{LS,L_\beta S_\beta }(k,p^{\prime})\
\tau_\beta\big(w_{\beta} (p^{\prime})\big)
Z^{\beta ,\alpha, T;J^\pi}_{L_\beta S_\beta ,L_\alpha S_\alpha}(p^{\prime},p)\,,
\end{eqnarray}
where $\alpha\in\{d,\Delta\}$, and $k$ denotes the momentum of the
incident photon. The argument $w_\beta$ of the propagator $\tau_\beta$
is the quasiparticle energy calculated on the assumption that the
corresponding spectator $b$ is on-shell:
\begin{equation}\label{Eq1_5}
w^2_\beta(p^{\prime})=W^2-2WE_b(p^{\prime})+M_b^2\,,
\end{equation}
with $E_b(p^{\prime})=\sqrt{p^{\prime\,2}+M_b^2}$. The spin $S$ of the
initial $\gamma d$ state in Eq.~(\ref{Eq1_4}) is a vector sum of the
deuteron spin $\vec{s}_d$ and the photon circular polarization vector
$\vec{\epsilon}_{\,\lambda}$ with components
$(\vec{\epsilon}_{\,\lambda})_\mu=-\delta_{-\mu\lambda}$.
The partial wave components of the driving terms
$Z^{\beta, \alpha,T ;J^\pi}_{L_\beta S_\beta ,L_\alpha S_\alpha}$
are obtained using the formalism developed, e.g. in Ref.~\cite{Rinat}.

In the present calculation we have included states with total angular
momentum up to $J_{max}=7$ of both parities with a maximum orbital
momentum $L_{max}=9$. Furthermore, since the dominating
$\Delta$-resonance term only enters states with total isospin
$T=1$, we neglect contributions of the $T=0$ part. Therefore, we omit in
the subsequent equations the isospin notation. As a
result, for each total spin and parity $J^\pi$ we have at most six
coupled one-dimensional
integral equations, two equations for the channel
$\alpha=d$ and four for $\alpha=\Delta$.

Our basic ingredient is a separable representation of the scattering
amplitudes in the $\pi N$ and $NN$ two-body subsystems. For $\pi N\to
\Delta\to \pi N$ we take
\begin{equation}\label{Eq1_6}
t_\Delta(\vec{q},\vec{q}^{\,\prime};z)=
g_\Delta(\vec{q}\,)\,\tau_\Delta(z)\,g^\dagger_\Delta(\vec{q}^{\,\prime})
\end{equation}
with
\begin{equation}\label{Eq1_6a}
\tau_\Delta(z)=\frac{1}{z-M^0_\Delta-\Sigma_\Delta(z)}\,,
\end{equation}
 where $M^0_\Delta$ denotes the bare $\Delta$ mass, and 
\begin{equation}
\Sigma_\Delta(z)=\frac14\sum_{m,m_\Delta}
\int\frac{d^3q}{(2\pi)^3}\,
\frac{\Big|\langle 
\frac12\,m| g_\Delta(\vec{q}\,)|\frac32\,m_\Delta\rangle\Big|^2}
{z-E_\pi(q)-E_N(q)+i\epsilon}
\end{equation}
the $\Delta\to \pi N$ self-energy.

\begin{figure}[ht]
\begin{center}
\resizebox{1.\textwidth}{!}{%
\includegraphics{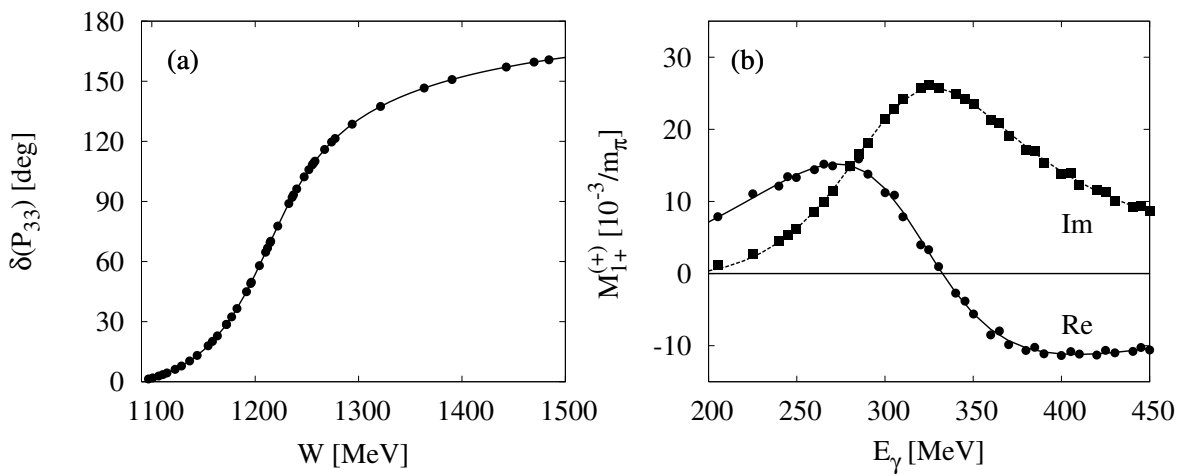}}
\caption{Panel (a): Present fit to the $P_{33}$ $\pi N$ phase shifts
  using Eqs.~(\ref{Eq1_6})-(\ref{Eq1_7}). The data are taken from
  the compilation in Ref.~\cite{Rowe}. Panel (b): The
  $M^{(+)}_{1+}$ multipole for $\gamma N\to \pi N$. Solid and dashed curves: our
  fit for real and imaginary parts, respectively. The full  circles and
  squares show the energy independent multipole
  analysis from Refs.~\cite{BerDon,Arndt1990}.}
\label{fig2}
\end{center}
\end{figure}

The vertex $g_\Delta$ is taken in the standard form (the isospin part
is omitted) with a monopole form factor 
\begin{equation}\label{Eq1_7}
g_\Delta(\vec{q}\,)=\frac{f_{\pi
N\Delta}}{m_\pi}\,(\vec{\sigma}_{\Delta N}\cdot\vec{q}\,)
\, \frac{\beta_\Delta^2}{\beta_\Delta^2 +q^2}
\sqrt{\frac{M_N}{2E_\pi(q) E_N(q)}}
\,,
\end{equation}
where $m_\pi$ denotes the pion mass and $\vec{\sigma}_{\Delta N}$ the
$N\to \Delta$ spin transition operator.
The parameters $M^0_\Delta$, $f_{\pi N\Delta}$ and $\beta_\Delta$ were
adjusted to the $\pi N$ phase shifts in the $P_{33}$ channel. The
resulting fit, presented in panel (a) of Fig.~\ref{fig2}, gives
$M^0_\Delta=1306$ MeV, $f^2_{\pi N\Delta}/4\pi=0.8113$, and
$\beta_\Delta=295$~MeV.

For the electromagnetic transition $\gamma N\to \Delta$ only the
dominant $M1$ part is 
taken into account. The corresponding vertex function was parametrized
in the form
\begin{equation}\label{Eq1_11}
g^\gamma_{\Delta}(z,\vec{k}\,)=e\frac{G^{M1}(z)}{2M_N}
\big(\vec{\sigma}_{\Delta N}\cdot(\vec{k}\times\vec{\epsilon_\lambda})\big) \,,
\end{equation}
with $e$ for the elementary charge, and the magnetic transition moment
\begin{equation}\label{Eq1_12a}
G^{M1}(z)=\mu _{\Delta }(z)\,e^{i\Phi _{\Delta } (z)}\,,
\end{equation}
with modulus $\mu _{\Delta }(z)$ and phase $\Phi _{\Delta } (z)$.

The off shell-behavior of the vertex (\ref{Eq1_11}) is determined by
the analytic continuation of the transition moment $G^{M1}(z)$ of
Eq.~(\ref{Eq1_12a}) into the complex plane of $z$. Below the
single-nucleon threshold we use
\begin{equation}\label{Eq1_12}
G^{M1}(z)=G^{M1}(m_\pi+M_N)\,,
\end{equation}
for $ \Re\,z< m_\pi+M_N$.
The approximation (\ref{Eq1_12}) obviously violates analyticity of the
amplitude. However, as the direct calculation shows, the subthreshold
region provides only a small fraction of the resulting cross section,
at least in the energy region not very close to the threshold, so that
this shortcoming of our model does not visibly affect the results.

Following Ref.~\cite{WilhArenh} we fit the energy dependence of
$\mu_{\Delta } (z)$ and $\Phi_{\Delta } (z)$ in such a way that the
resulting $\gamma N\to\Delta\to \pi N$ amplitude
\begin{equation}\label{Eq1_13}
t_\Delta^\gamma(\vec{k},\vec{q};z)=g_{\Delta}^\dagger  (\vec{q}\,) \,
\tau_\Delta(z)\, g^\gamma_{\Delta}(z,\vec{k}\,)
\end{equation}
reproduces the isovector magnetic amplitude $M^{(+)}_{1+}$ in the
energy region from threshold up to 450~MeV (panel (b) in
Fig.~\ref{fig2}). Thus, we do not treat the background terms 
(crossed nucleon pole and $\omega$-exchange) exactly, but their
contribution is effectively included via adjustment of the ansatz
in Eqs.~(\ref{Eq1_11})-(\ref{Eq1_13}) to the data of $M^{(+)}_{1+}$. The
magnitude $\mu_\Delta(z)$ and the phase $\Phi _{\Delta } (z)$ in
Eq.~(\ref{Eq1_12a}) are parametrized as
\begin{equation}\label{Eq1_14}
\mu_{\Delta }(z)=\sum\limits_{n=0}^{4}C_n\left(\frac{z}{M_\Delta}\right)^n\,,\quad
\Phi_\Delta(z)=\sum\limits_{n=0}^{4}D_n\left(\frac{z}{M_\Delta}\right)^n\,,
\end{equation}
with $M_\Delta=1232$ MeV. The constants $C_n$ and $D_n$ resulting from
the fit in Fig.~\ref{fig2} are collected in Table \ref{tab1}.

\begin{table}[h]
\caption{Listing of constants $C_n$ and $D_n$ of the parametrizations
  in Eq.~(\ref{Eq1_14}).}
\begin{ruledtabular}
\begin{tabular}{r|r|r|r|r|r}
$n$ & 0  & 1 & 2 & 3 & 4  \\
\colrule
$C_n$ &  37.848 & $-29.789$ & $-19.951$  & \ 12.261 & 4.4393 \\
$D_n$ & $-12.901$ &  19.163 &  4.4974 & $-14.938$ & 4.2943 \\
\end{tabular}
\end{ruledtabular}
\label{tab1}
\end{table}

In the $NN$ sector only the $s$-waves $^3S_1$ and $^1S_0$ are
taken into account, neglecting the contribution of the tensor
component $^3D_1$. For the $s$-wave interactions we use the rank-one
separable parametrization of the Paris potential from
Ref.~\cite{Zankel}:
\begin{equation}
v_d^{(s)}(\vec{q},\vec{q}^{\,\prime})=-g_d^{(s)}(q)\,g_d^{(s)}(q^\prime\,)
\end{equation}
with
\begin{equation}
g_d^{(s)}(q)=(2\pi)^{3/2}\sum\limits_{n=1}^6\frac{C^{(s)}_n}{q^2+(\beta_n^{(s)})^2}\,,
\end{equation}
where the index $s$ refers to singlet or triplet states.
The parameters $C^{(s)}_n$ and $\beta_n^{(s)}$ are listed in \cite{Zankel}.

The coupled integral equations in Eq.~(\ref{Eq1_4}) were solved using the
matrix inversion method. To overcome the problem of singularities we
used the well known procedure in which the integration contour is
shifted from the real axes to the fourth quadrant of the complex
$p^{\prime}$ plane. This technique is quite well known (e.g., see
Ref.~\cite{Aaron}), and there is no need to describe it here in
detail. Some formal aspects related to the relativistic kinematics
were considered in Ref.~\cite{FiAr2001}.

\begin{figure}[ht]
\begin{center}
\resizebox{0.5\textwidth}{!}{%
\includegraphics{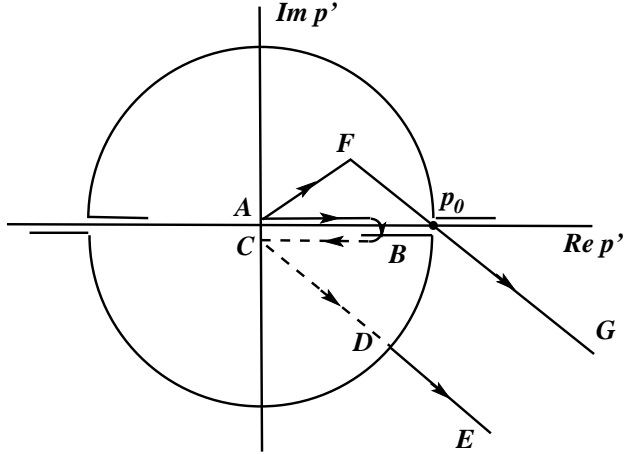}}
\caption{Integration contours and configuration of the cuts in the
  complex  $p^{\prime}$ plane. The dashed line shows the part of the
  contour on the second sheet.}
\label{fig3}
\end{center}
\end{figure}

Here we would like to
comment only on some details concerning the treatment of the
singularities 
of the driving terms
$Z^{\beta, \alpha ,T;J^{\pi}}_{L_\beta S_\beta ,L_\alpha S_\alpha}$
having the configuration shown in Fig.~\ref{fig3}. It is known, that in
order to find the $X^\alpha$-matrix at real momenta, one has 
to perform a continuation of the driving terms $Z^{\beta, \alpha
  ,T;J^{\pi}}_{L_\beta S_\beta ,L_\alpha S_\alpha}$ in
Eq.~(\ref{Eq1_4}) onto the second Riemann sheet (the part $BCD$ of the
integration contour in Fig.~\ref{fig3}). However, in this case the
driving term $Z^{\beta, \alpha ,T;J^\pi
  }_{L_\beta S_\beta ,L_\alpha S_\alpha}(p^{\prime},p)$ behaves
near $p^{\prime}=0$ 
as $(1/p^{\prime})^{L+1}$.
As a result the integrand in Eq.~(\ref{Eq1_4})
strongly diverges at the origin and one finds large
contributions near the point $C$. However, the resulting large
contributions coming from the intervals $BC$ and
$CD$ essentially cancel each other thus leading to a significant loss of numerical
accuracy via a small difference of two large numbers. This problem was
also mentioned in Ref.~\cite{CahSlo},
where the break-up reaction $nd\to nnp$ was studied.
In Ref.~\cite{CahSlo} the three-body equations were solved only for low
values of $L$ ($L\leq 3$), whereas for higher $L$ only
the first order approximation or the
inhomogeneous term was considered.

In the present case, we use another integration contour (the polygonal
curve $AFG$ in Fig.~\ref{fig3}). The driving terms are always calculated on
the first Riemann sheet, and the integration does not pose any
numerical
problem. A certain disadvantage of this method lies in the fact that
the position of the momentum $p_0$, where the contour is squeezed
between the logarithmic cuts (see Fig.~\ref{fig3}), depends on the
value of the on-shell momentum $p$ in Eq.~(\ref{Eq1_4}). For this reason
one has to solve the set of  Eq.~(\ref{Eq1_4}) separately
for each value of $p$ of the chosen mesh.

After inversion of the system in Eq.~(\ref{Eq1_4}) and the
determination of the partial wave
amplitudes $X^{\alpha, T;J^\pi}_{LS,L_\alpha S_\alpha}(k,p;W)$,
one obtains the corresponding channel
amplitudes $X^{\alpha}_{\lambda M_d\,S_\alpha M_{S_\alpha}}
(\vec{k},\vec{p};W)$ from Eq.~(\ref{Eq1_16}), from which, finally, the
reaction amplitude $T_{\lambda M_d\,m_1m_2}$ (see
Eq.~(\ref{T-matrix})) as function
of the momenta of the final particles $\vec{q}_\pi$, $\vec{p}_1$, and
$\vec{p}_2$ follows
\begin{eqnarray}\label{Eq1_17}
T_{\lambda M_d\,m_1m_2}(k,\vec{p}_1,\,\vec{p}_2,\,\vec{q}_\pi)&=&
\sum\limits_{M_{d}^{\prime} }X^d_{\lambda M_d\,1 M_{d}^{\prime} }(k,\vec{q}_\pi;W)
\Big(\frac{1}{2}m_1\,\frac{1}{2}m_2\Big|1 M_{d}^{\prime} \Big)\tau_d^{(1)}
\big(w_d(q_\pi)\big)\,g_d^{(1)}(q_{NN})
\nonumber \\
&+&\Big[\sum\limits_{m_\Delta}
\sum\limits_{S_\Delta, M_{\Delta}} \sqrt{\frac{2}{3}}\,
X^\Delta_{\lambda M_d\,S_\Delta M_{\Delta}}(k,\vec{p}_1;W)
\Big(\frac{1}{2}m_2\,\frac32 m_\Delta\Big|S_\Delta M_{\Delta}\Big)
\nonumber\\
&&\times
\tau_\Delta\big(w_\Delta(p_1)\big)\,
\langle \frac32 m_\Delta|g_\Delta(\vec{q}_{\pi N_2})
|\frac12 m_1\rangle
-(1\leftrightarrow 2)\Big]
~.
\end{eqnarray}
Here the vector $q_{\pi N_i}$, $i=1,2$, denotes the relative momentum in the
subsystem $\pi N_i$, and $q_{NN}$ the relative momentum of the two final
nucleons.
As is mentioned above, in the present calculation we took into account
only configurations with total isospin $T=1$, since those with
$T=0$ do not contain the dominant $N\Delta$-configuration. Therefore,
in the first term on the rhs of Eq.~(\ref{Eq1_17}) only the two-nucleon states
with total spin $S_d=1$ contribute.

Using the amplitude of Eq.~(\ref{Eq1_17}), the
fully exclusive differential cross section for the present reaction in the
overall center-of-mass frame is given in terms of the $T$-matrix
(Eq.~(\ref{Eq1_17}))
\begin{equation}\label{Eq1_18}
\frac{d\sigma}{dq_\pi\,d\Omega_\pi\,d\Omega^*_{NN}}=\frac{1}{(2\pi)^5}
\frac{E_d\,E_p\,E_n\,q_\pi^2p^*_{NN}}{2W\omega_\gamma\,\omega_{NN}}
\frac{1}{6}\sum_{\lambda, M_d, m_1, m_2}|T_{\lambda M_d\,m_1m_2}
(k,\vec{p}_1,\vec{p}_2,\vec{q}_\pi)|^2\,,
\end{equation}
where $E_d$, $E_p$, $E_n$, and $\omega_\gamma$ denote the total
energies of the corresponding particles, and $\omega_{NN}$ the
invariant $NN$ energy. The nucleon momentum in the $np$ center-of-mass
system is denoted by $p^*_{NN}$ and its spherical angle by
$\Omega^*_{NN}$. 

\section{Results and discussion}\label{SectResults}

Before applying the formalism to incoherent pion photoproduction,
  we first test our model by considering the inelastic scattering of pions
  on a deuteron. The corresponding equations are obtained from
  (\ref{Eq1_1}) by the replacement of the driving term according to
$Z^{\gamma d,\Delta}\to Z^{\pi d,\Delta}$. The method of inversion of
the corresponding three-particle equations remains of course the
same.  

\begin{figure}[ht]
\begin{center}
\resizebox{0.8\textwidth}{!}{%
\includegraphics{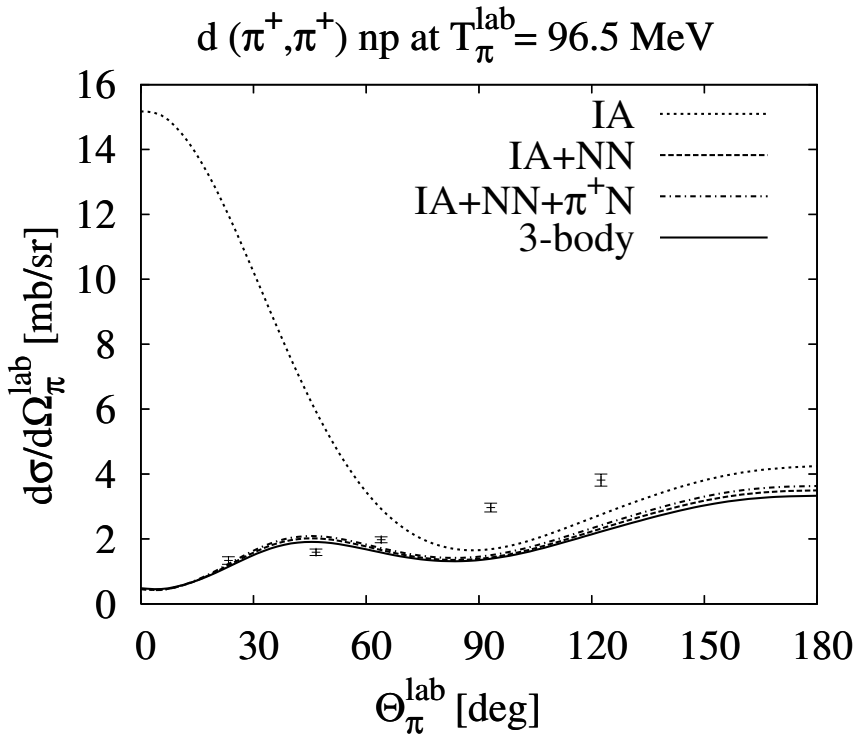}}
\caption{Differential cross section of the reaction
  $\pi^+ d\to\pi^+ np$ for an incident pion lab kinetic energy of 96.5
  MeV. The dotted curve is the result of the impulse 
  approximation (IA). Dashed and dash-dotted curves are obtained with
  first order $np$ and in addition $\pi^+ N$ rescattering
  contributions. Solid curve: three-body calculation. The data are
  from Ref.~\cite{Khandak}.} 
\label{fig3a}
\end{center}
\end{figure}
 The results are presented in Fig.~\ref{fig3a}. As already noted
  in~\cite{Garc1}, the significant influence of the final state
  interaction on the magnitude and shape of the differential cross
  section is due to the orthogonality of the wave functions of the initial
  and the final $np$ states. In particular, this effect leads to a
  substantial suppression of the plane wave cross section (IA) at forward
  angles, and, as a result, to a general agreement with the
  data~\cite{Khandak} in the forward direction.   

In the region around $\theta_{lab}=90^\circ$ our cross section
underestimates the experimental results and turns out to be lower than
the theoretical cross section obtained in Ref.~\cite{Garc1}. The
latter deviation is apparently caused by a disagreement seen already
between the IA results. Namely, our plane wave cross section is almost
twice as small  in this region as that obtained in~\cite{Garc1}. At
the same time, the FSI effects predicted by our model and in
Ref.~\cite{Garc1} are in reasonable agreement.  

As the analysis of the curves in Fig.~\ref{fig3a} shows, the FSI
effect is almost completely reproduced by including only the
first-order rescattering contributions. In particular, this concerns the
$np$ interaction  whereas pion rescattering gives only a small
addition at the level of 0.5-1~\%. Inclusion of the remaining
higher order terms within the three-particle model does not lead to
any noticeable change in the cross section in the entire range of 
pion angles. Since the dynamics of the process $\gamma d\to \pi^0 np$
associated with FSI is essentially similar to that of the inelastic
$\pi^+d$  scattering, it is reasonable to expect that qualitatively
the same picture will be observed in the incoherent pion
photoproduction. 

\begin{figure}[!h]
\begin{center}
\resizebox{1.\textwidth}{!}{%
\includegraphics{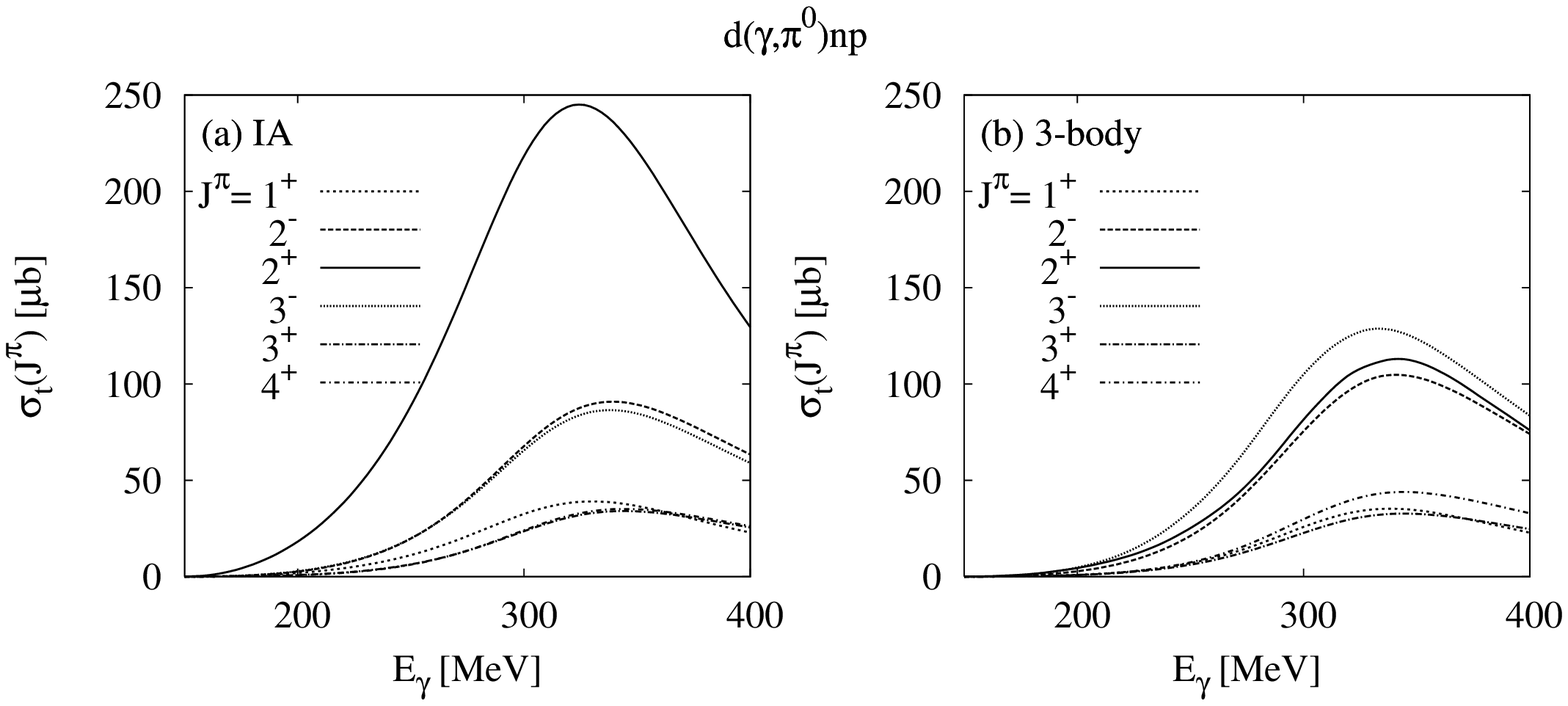}}
\caption{ Contributions of various partial waves $J^\pi$ to the total cross
section $\gamma d\to\pi^0 np$: Panel (a): impulse approximation: Panel
(b):  three-body calculation. }
\label{fig4}
\end{center}
\end{figure}

Turning now to the reaction $\gamma d\to \pi^0 np$ we start the
discussion by considering the role of different partial waves in the
total cross section as is shown in Fig.~\ref{fig4} for the IA (panel
(a)) and the full three-body calculation (panel (b)). Similar to the
coherent photoproduction $\gamma d\to\pi^0 d$~\cite{WA95} the largest
contribution in the incoherent reaction comes from the $2^+$ wave,
which predominantly is an $M1$ transition, leading to the production
of  pions with angular momentum $l_\pi=1$ with respect to the $np$ 
system. This partial wave alone contributes almost $45~\%$ to the
total IA cross section in the $\Delta$ region. The next important
partial waves are $2^-$ and $3^-$ generating basically pions with
$l_\pi=2$. The other partial waves give much smaller contributions to
the total cross section.  

\begin{figure}[!h]
\begin{center}
\resizebox{1.\textwidth}{!}{%
\includegraphics{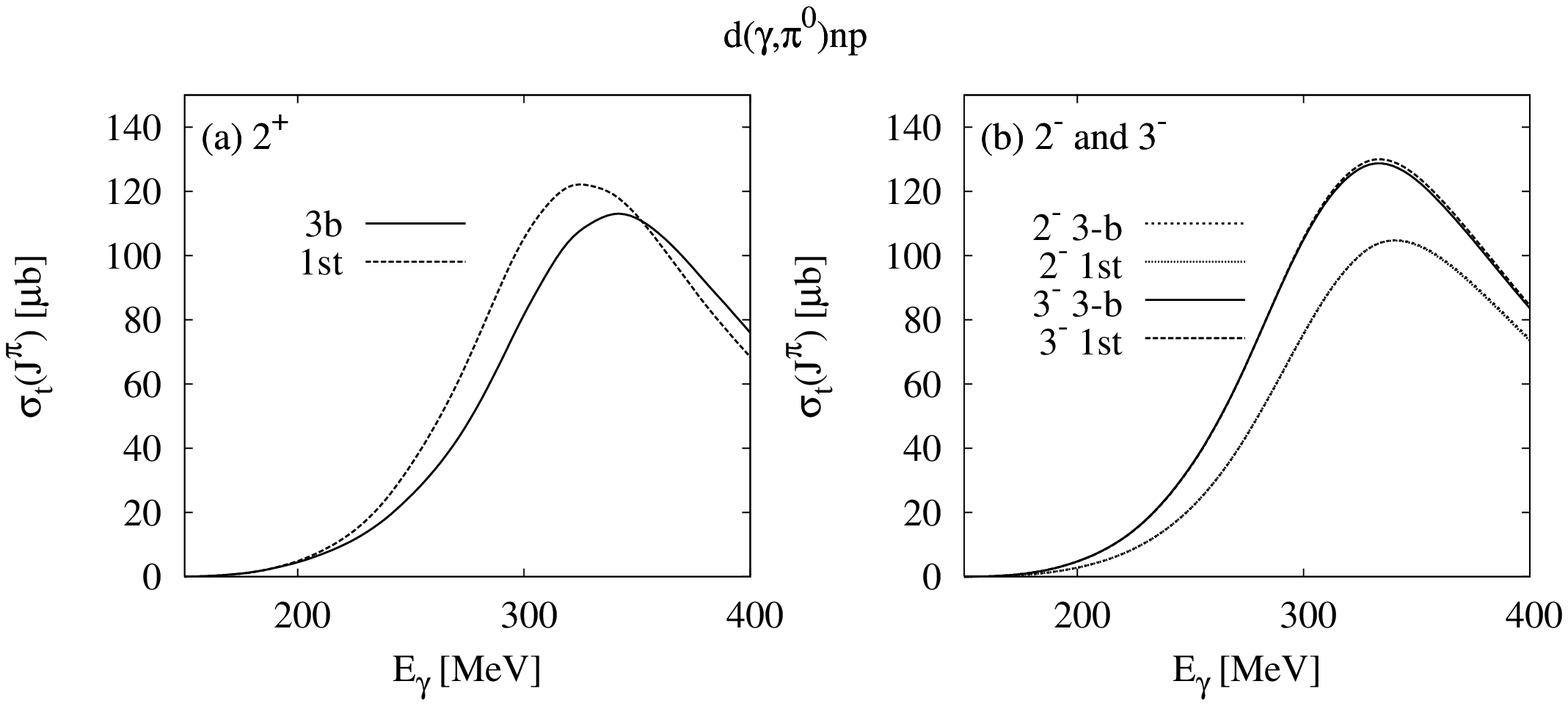}}
\caption{ Comparison of the first order rescattering calculation (1st) with
  the three-body one (3-b) for the most important partial waves: 
  Panel (a): $2^+$; Panel (b): $2^-$ and $3^-$.}
\label{fig4a}
\end{center}
\end{figure}

Inclusion of FSI (see panel (b) of Fig.~\ref{fig4}) leads to a visible
decrease of $\sigma (2^+)$ by a
factor 2-3, whereas $\sigma (3^-)$ (predominantly $M2$) is considerably
enhanced and becomes even slightly larger than $\sigma (2^+)$. Next in importance
is $\sigma (2^-)$, which is also increased by FSI by about 10~\%. The
contributions of the higher partial waves is still quite
insignificant. Most of the FSI effects are already contained in the
first order rescattering contribution except for the dominant
$2^+$-partial wave as is demonstrated in Fig.~\ref{fig4a} where the
higher orders still give a significant contribution. It is worth
noting that the state $2^+$ corresponds to the $s$-wave $\Delta N$
configuration $^5S_2$. In this respect, the importance of three-body
effects in this wave agrees with our naive expectation that the absence of
the centrifugal barrier in the $^5S_2$ state leads to a significant
overlap of the potentials in all three two-body subsystems.

\begin{figure}[!h]
\begin{center}
\resizebox{.8\textwidth}{!}{%
\includegraphics{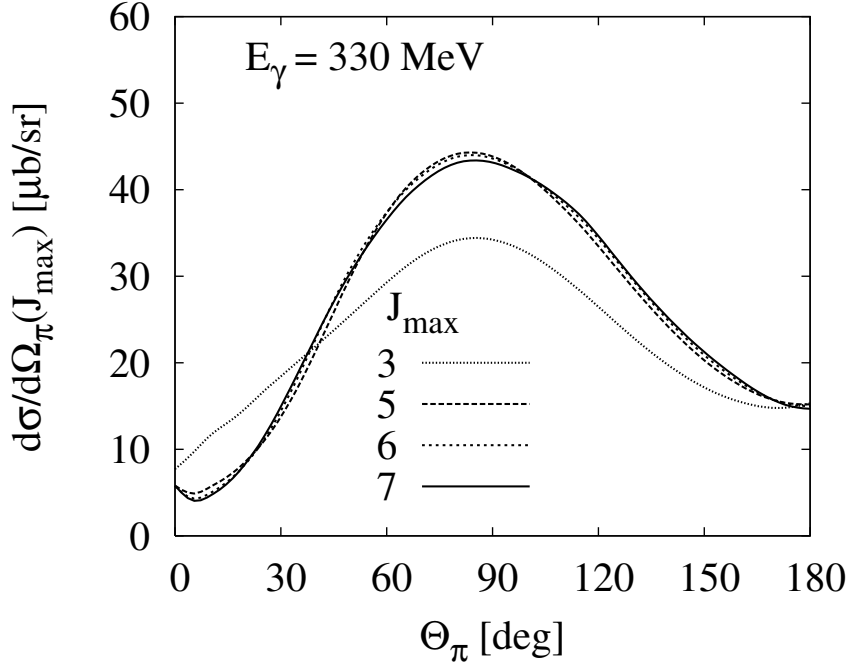}}
\caption{ Differential
cross section $\gamma d\to\pi^0 np$  in the
center-of-mass frame at $E_\gamma=330$ MeV for various values of the
maximum total   angular momentum $J_{max}$ of the partial wave
expansion. }
\label{fig4b}
\end{center}
\end{figure}

As already mentioned, the
corresponding partial wave series was cut-off at $J_{max}=7$ and
$L_{max}=9$. In order to demonstrate the good convergence for this
value of $J_{max}$ we show in Fig.~\ref{fig4b} the semi-exclusive differential
cross section with respect to the final pion as function of a few
lower $J_{max}$-values.  Similar to the results for the elastic
pion-deuteron scattering of Ref.~\cite{RinatThomas}, this approximation
provides a satisfactory convergence in the $\Delta$-resonance
region as one
can see: changing $J_{max}$ from 5 to 7 has already quite a small
effect, so that the choice $J_{max}=7$ appears to be acceptable.

\begin{figure}[!h]
\begin{center}
\resizebox{1.\textwidth}{!}{%
\includegraphics{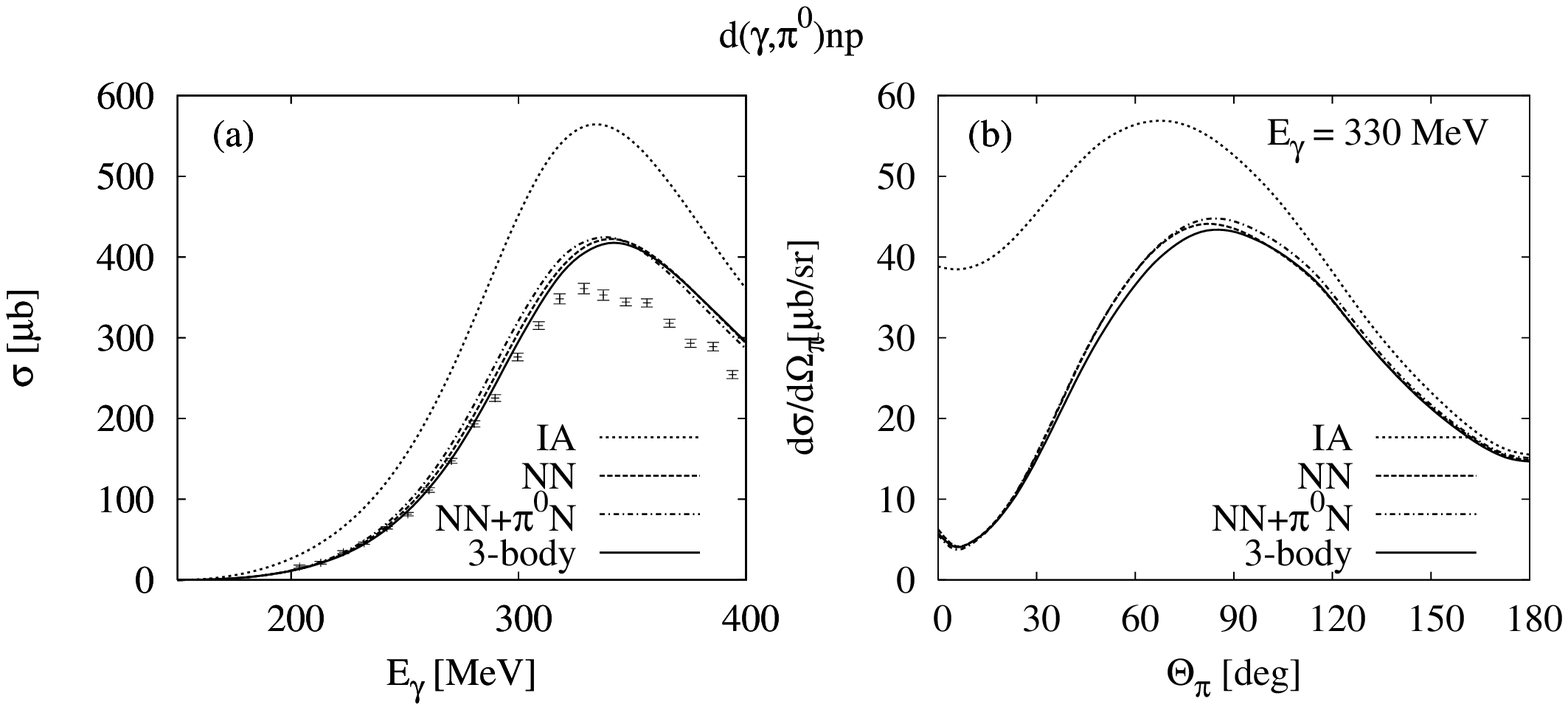}}
\caption{Total (panel (a)) and differential (panel (b)) cross sections of the reaction
  $\gamma d\to\pi^0 np$. Dotted curves: impulse
  approximation (IA) (the inhomogeneous $Z^{\gamma d,\alpha}$ terms in
  Fig.~\protect\ref{fig1}). Dashed and dash-dot curves include
  first order $np$ and in addition $\pi^0 N$ rescattering
  contributions, respectively. Solid curves: full three-body calculation. 
The data are from Ref.~\cite{Krusche}.}
\label{fig5}
\end{center}
\end{figure}

In Fig.~\ref{fig5} we show in panel (a) the total cross
section as function of the incident photon energy and in panel (b) the
differential cross section at $E_\gamma=330$~MeV, calculated in the
present three-body model. These results can be compared 
with our previous calculation in Ref.~\cite{FixArenh}. In the latter
case the single nucleon amplitude $t(\gamma N\to \pi N)$ was taken
from the MAID analysis~\cite{MAID} and the inclusion of FSI was reduced to
the first order contributions (i.e., to $np$ and $\pi N$
rescatterings in the final state). The present result, which is
obtained in a somewhat oversimplified model for $\gamma N\to \pi N$
(pure resonance ansatz for the $M^{(+)}_{1+}$ multipole, and neglect of
the tensor component of the deuteron wave function) agrees quite well
with those of Ref.~\cite{FixArenh}. Fig.~\ref{fig5} clearly shows that
the multiple scattering corrections do not visibly change the reaction
dynamics. Inclusion of only the first order corrections, i.e.\ $np$
and $\pi N$ rescatterings in the final state, turns out to be
sufficient. 

In the same Fig.~\ref{fig5} (panel (a)) we compare our results with
experimental total cross section data from
Ref.~\cite{Krusche}. Whereas in the region 
below the $\Delta(1232)$ peak the agreement is satisfactory,  the
theory is too high in the region near the maximum ($E_\gamma\ge
300$~MeV). This discrepancy was already discussed in
Refs.~\cite{Darwish,FixArenh,Levchuk}. In particular it was
conjectured in \cite{FixArenh} that the difference may come from the
neglect of the $\Delta N$ interaction which might lead to a broadening
of the $\Delta$ resonance due to additional inelasticities. The
present calculation, which effectively takes into account the $\Delta
N$ interaction, shows, that this effect is negligible in the
incoherent reaction and cannot explain the discrepancy. 

 There is still the unresolved question about the importance of true
  pion absorption, which is not taken into account by the present
  model. We however assume that its role in our reaction is not
  significant for the following reasons. Firstly, the two-nucleon absorption of pions is
  effective only in the region of small internucleon distances, which
  are not important in the breakup process. Secondly, the calculations
  performed by us in the framework of the so-called bound state
  picture (BSP), in which one of the nucleons is represented as a
  bound $\pi N$ state with the quantum numbers $P_{11}$
  \cite{Aaron1969}, give a correction to the total cross section of
  only about 0.5~$\%$. 
It is worth noting, that the BSP-based approach is not entirely
correct, since in fact it treats the nucleons in the intermediate $NN$
states as distinguishable particles, and today various sophisticated
methods have been developed to incorporate the $NN$ channel into the
three-particle $\pi NN$ equations. However, one can hardly expect that
even a correct treatment of the $NN$ states will dramatically
increase their role in our reaction. Our assumption about the
insignificance of true pion absorption is also in accord with the
results of Ref.~\cite{Garc2} where a similar conclusion was reached for
the $\pi d$ inelastic scattering.

\section{Conclusion}\label{Conclusion}

In this paper we have presented a calculation of total and differential
cross sections for the incoherent reaction $\gamma d\to \pi^0 np$ in
the energy region from threshold up to the $\Delta$ resonance. The
calculation is based on a three-body model for the inclusion of the final $\pi NN$
interaction. Although we use some simplifications (non-relativistic
three-body equations, neglect of the deuteron $d$-wave), the most
significant features of the process  are preserved, including the
importance of the $M1$ multipole transition $\gamma N\to\Delta$ and
the dominance of the $\Delta$ resonance in $\pi N$ scattering.

The results show that the corrections due to
multiple scattering are quite insignificant in the major part of the
kinematical region. As already mentioned in the introduction, the major
importance of FSI in the $\pi^0np$ channel is related to the
orthogonality of the initial and the final $np$ wave functions. This
effect is taken into account essentially already by the first order FSI
contributions as can be seen from the fact that inclusion of the $np$
rescattering (dashed curve on the right panel of Fig.~\ref{fig5})
leads to a significant decrease of the differential cross section at
very forward pion angles, that is, in the
region where the momentum transferred to the nucleon system is minimal.

According to our results, the full three-body calculation changes the
cross section compared to the first-order rescatterings only by about
$1-2~\%$ in the $\Delta$ resonance region. Since the multiple
scattering corrections are insignificant their inclusion cannot
explain the existing deviation between the theoretical and
experimental results. The theory still
visibly overestimates the data, as is shown in panel (a) of Fig.~\ref{fig5}. The
problem concerning the difficulties in describing the photoproduction of
$\pi^0$ on a deuteron in the first resonance region was also addressed
in Ref.~\cite{Siodlaczek}. In this work the authors had analysed the
inclusive cross section $\gamma d\to \pi^0 X$ with $X$ being either a
deuteron or a neutron-proton scattering state.

As is shown in
Ref.~\cite{Kolybasov} using the closure approximation for the final
two-nucleon state, the sum of both cross sections should be equal to
the sum of the free-nucleon cross sections, folded with the nucleon
momentum distribution in a deuteron. The latter is approximately equal
to the cross section $\sigma_{IA}$ of $\sigma(\gamma d\to \pi^0 np)$
calculated in the spectator model. However, as the calculation in
Ref.~\cite{Siodlaczek} shows, $\sigma_{IA}$
overestimates by about 15~$\%$ the experimental total cross section
for $\gamma d\to \pi^0 X$. Since the free proton cross section is well
known the natural conclusion would be that the free neutron cross
section is overestimated by the existing multipole analyses (in
Ref.~\cite{Siodlaczek} the MAID~\cite{MAID} and SAID~\cite{SAID}
analyses are considered).
In order to bring the theory into agreement with the data of
\cite{Krusche,Siodlaczek} the theoretical neutron cross section has to
be decreased by about $25~\%$. Such a strong isospin dependence of
the elementary amplitude can hardly be explained within the existing
models for $\gamma N\to \pi N$. According to these models the reaction
is strongly dominated by the $\Delta(1232)$ resonance so that the
proton and neutron cross sections are nearly equal. Thus the question
about the source of the discrepancy between
theoretical predictions and data in the $\Delta(1232)$ region
remains open.

\begin{acknowledgments}
This work was supported by the Deutsche Forschungsgemeinschaft
(Collaborative Research Center 1044).
A.F. acknowledges additional support from the Tomsk Polytechnic
University Competitiveness Enhancement Program. 
\end{acknowledgments}

\end{document}